# EFFECT OF TIME AND THERMO-MECHANICAL COUPLINGS ON POLYMERS

*Pankaj Yadav #1 [a], André Chrysochoos #2 [a], Olivier Arnould #3 [a], Sandrine Bardet #4 [a]*

[a]LMGC – Laboratoire de Méchanique et Génie Civil, CNRS UMR-5508, Université de Montpellier, Montpellier, France, pankaj.yadav@etu.umontpellier.fr #1, andre.chrysochoos@umontpellier.fr #2, olivier.arnould@umontpellier.fr #3, sandrine.bardet@umontpellier.fr #4

Analysis of the thermo-mechanical behaviour of polymers has been and still is the subject of many rheological studies both experimentally and theoretically. For small deformations, the modelling framework retained by rheologists is often of linear visco-elasticity, which led to the definition of complex modules and used to identify the glass transition temperature as the so called rule of time-temperature superposition. In this context, the effects of time are almost unanimously associated with viscous effects. It has also been observed that the dissipative effects associated with viscous effects are often very small compared to the coupling of sources indicating a high sensitivity of polymeric materials to temperature variations.

This work is mainly focused on establishing the exact role of coupling effects, which also induce the effect of time. Using traditional experimental methods of visco-analysis (DMTA) and via an energy analysis of the behaviour, the goal of the thesis is to try to restate the time-temperature equivalence rule under the Thermodynamics of Irreversible Processes, taking into account the dissipative effects and coupling induced process deformation.

## 1 Introduction

Polymeric materials are widely known for their high viscoelasticity, which signifies wide dependence of their mechanical properties on the span or frequency of the applied stress which is very important in exemplary engineering applications. This is one of the reason why it was desired to formulate a method which could be capable of predicting the viscoelastic behaviour of polymeric materials over time scales that extend as long as the working life of the material or to gather behaviour at easily handlable frequencies in laboratories. Since it was normally difficult or lets say was impossible to operate such a lengthy relaxation or creep or frequency sweep over a wide range of loading frequencies, likely methods were used to figure the time dependency of viscoelastic behaviour of the polymeric materials. In early 1940's, Leaderman[1] (cited in[2]) was the first who recognised the aftereffects of the similarity between creep curves measured at closely separated temperatures. This striking similarity between viscoelastic parameters when measured at closely spaced temperatures, was found to be a common occurence and is not confined to a few relaxations. Specifically, the procedure that was used for superimposing the curves and to predict this behaviour over a wide span of frequency has come to be known as time-temperature superposition principle.

The approach of time-temperature shift is expressed as mentioned in equation 1, with J(t,T) represents the creep compliance as a function of creep time t at the temperature T. The creep functions at temperature T and $T_0$ should be connected by equation(1) and equation (2) is a function of the temperatures T and $T_0$ only[2].

where
$$J(t, T) = J(a.t, T_0) \quad (1)$$
$$a = a(T, T_0) \quad (2)$$

In general, the characterization of material behaviour under cyclic loading conditions involves two closely related aspects, namely mechanical and thermal, that can be grouped into one general thermo-mechanical framework. The material often shows stabilized hysteretic responses as the results of cyclic loading. There are several techniques which are always advantageous in characterising the energy nature of such hysteresis loops. Generally, dissipative mechanisms are responsible for the mechanical energy lost in such hysteresis loops, reflecting irreversible material degradation [3]. Previous work of researchers have found that thermodynamic analysis of the cyclic responses revealed that such hysteresis areas may not only be induced by intrinsic dissipation but also by internal energy variations in the form of stored energy and/or by strong thermo-mechanical coupling effects associated with the heat diffusion [3-6].

## 2 Experimental section

The technique used to carry out the viscoelastic measurements on PMMA and PS was the traditional equipment known as DMTA. The DMA machine used for these measurements is 242e Artemis by NETZSCH. Functional representation of the machine is shown by Figure 1.

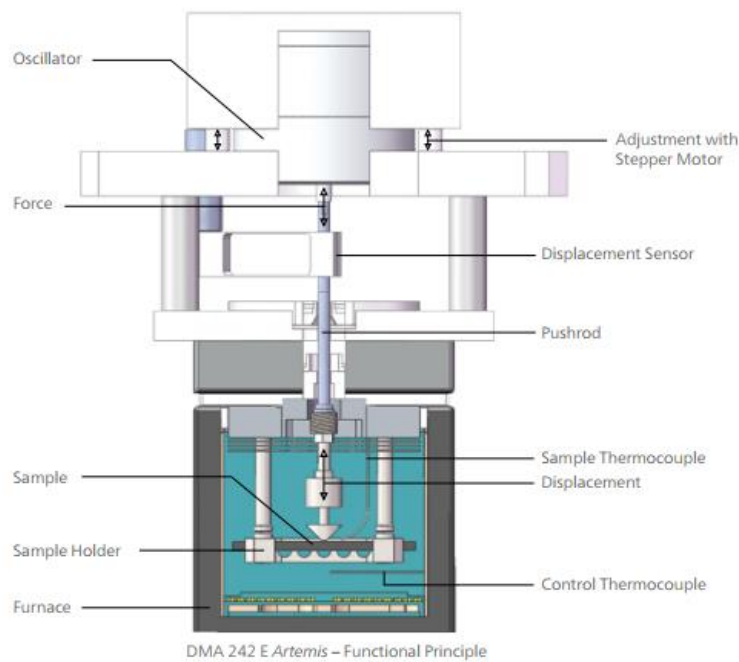

FIGURE 1 – Representation of functional principle of DMA 242e Artemis

The samples used for visco-elastic measurements of PS by DMA were rectangular bars of cross-section 1.2 x 4 mm$^2$ and the tension mode was employed. Visco-elastic studies were carried out by the DMA shown in Figure 1 to obtain isothermal dynamic moduli data E'(storage modulus) and E''(loss modulus). The experimental frequencies ranged from 0.01 to 10 Hz for each temperature. The range of temperature used was 50°C to 110°C for PS. Temperature regulation was done by using $N_2$ gas in order to avoid any structural modification. In all cases we have ensured the linearity of the stress response with respect to the strain amplitude in order to remain within the linear visco-elastic region. New sample was used for each temperature as using the same sample for all the temperatures have intrinsically a larger experimental error [26]. The desired temperatures were achieved with the increment of 2 K/minute followed by an equilibrium time of 1800 seconds provided to the sample after reaching the desired temperature to equalize the temperature throughout the sample before starting the cycle of measurements which lasts about 330 minutes.

The glass transition temperature ($T_g$) was determined using the same equipment by employing the TMA mode in the temperature range from 0°C to 150°C.

## 3 Results

The raw data was recorded for each sinusoidal signal directly from the DMA machine in the form of ".csv" to verify if the response is monochromatic. To verify the response, FFT (Fast Fourier Transform) technique was employed on the raw data using multi-paradigm numerical modelling environment of MATLAB. A Fast Fourier transform algorithm computes the discrete Fourier transform (DFT) of a sequence, or its inverse (IFFT). Fourier analysis converts a signal from its original domain (often time or space) to a representation in the frequency domain and vice versa. FFT of the sinusoidal signal at 0.1 Hz frequency is shown in Figure 2.

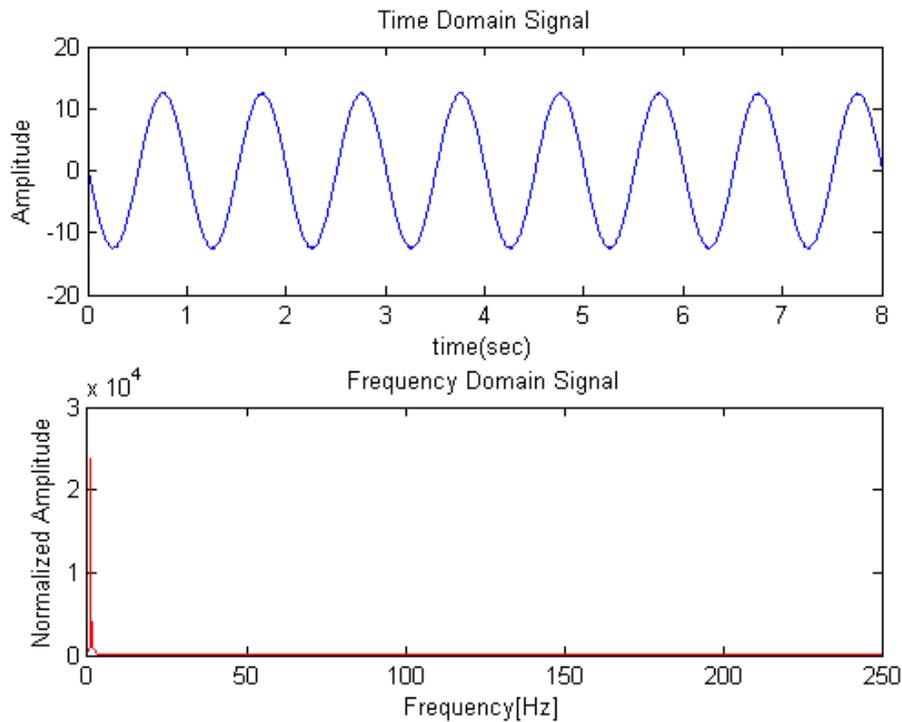

FIGURE 2 - Representation of cyclic load before and after Fast Fourier Transform

The viscoelastic results obtained from frequency sweeps of PS are shown in Figure 3(a) and (b). (b)

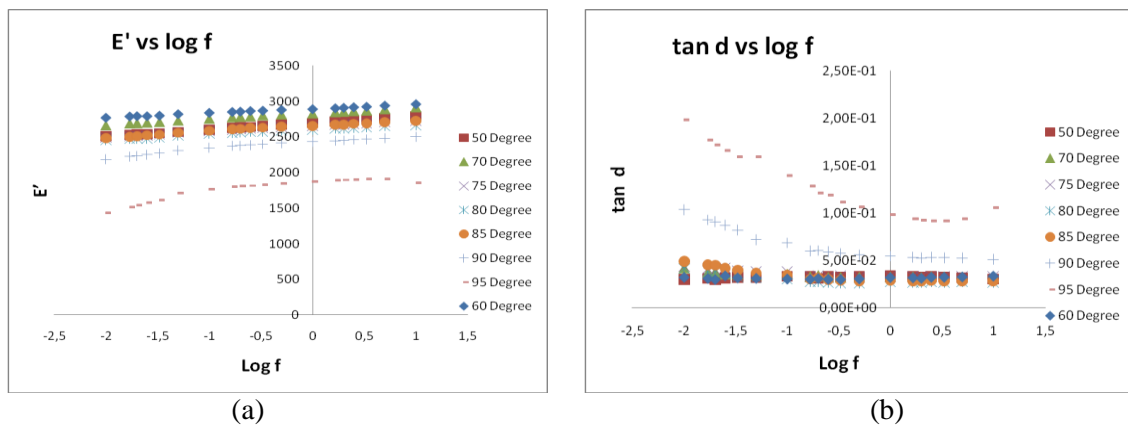

(a)             (b)

FIGURE 3 – Frequency sweep results of PS (a) storage modulus and (b) loss tangent

The dynamic mechanical spectra have been represented as the values of E' and tan δ versus logarithmic frequency for PS in Figure 3(a) and (b) for the temperature range from 50ºC-110ºC.

Later, the frequency sweep curves at different temperatures were superimposed by using manual fitting to make a master curve as manual fitting was supposed to give the best fitting along with mathematical fitting [7]. The master curve made is shown in Figure 4 for (a) storage modulus and (b) loss tangent.

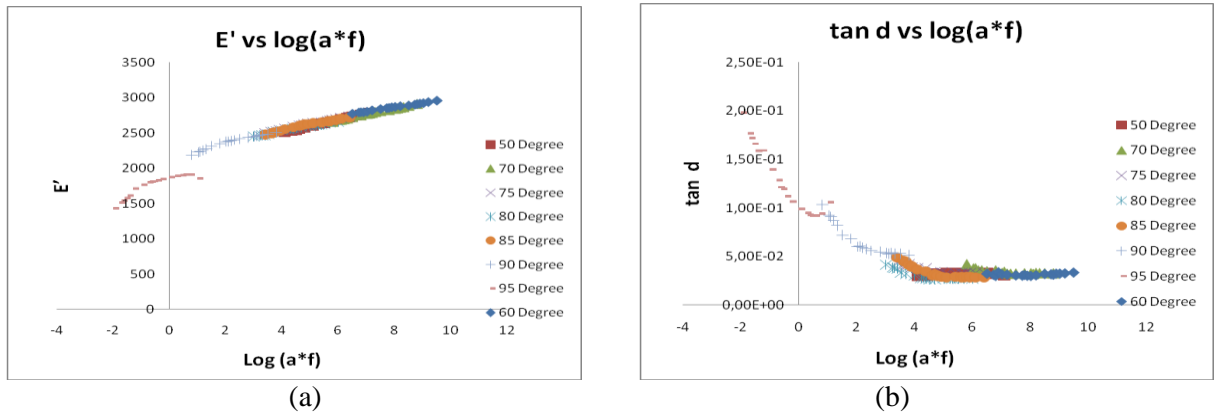

FIGURE 4 – Master curves plotted for (a) storage modulus and (b) loss tangent with horizontal shifts

Initially, the curves were displaced along frequency scale alone and was observed during construction of master curve that superposition by displacement along frequency scale alone cannot yield a master curve for loss tangent spectrum which is in accordance to Cavaille et al [8], whereas, he also found a satisfactory superposition of storage modulus by displacement along frequency scale alone.

Later, the horizontal shift was supplemented by vertical shifts of the modulus due to the free volume concept of polymers and was found that a very well master curve was obtained. According to the Rouse model [9], the modulus is multiplied by a factor $b_T$ and is given as,

$$b_T = T_0\rho_0 / T\rho \qquad (3)$$

After supplementing horizontal shift with vertical shifts, results are shown in Figure 5.

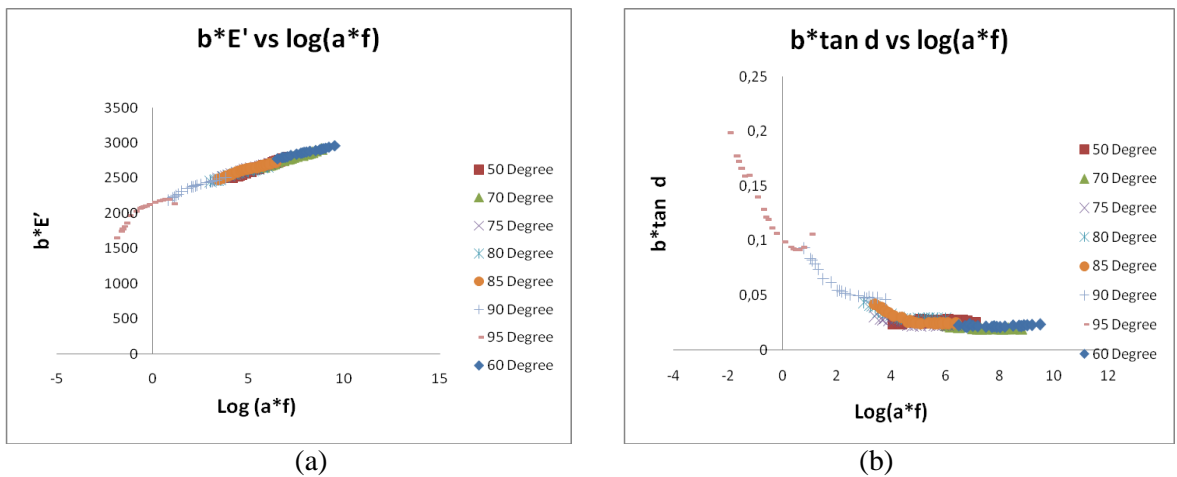

FIGURE 5 - Master curves plotted for (a) storage modulus and (b) loss tangent with both shifts

# 4 Conclusion

In this work, implementation of time-temperature superposition was tried on the visco-elastic properties of the amorphous polymer PS (Polystyrene) by using manual fitting method. It was observed that implementation of time-temperature superposition works in a well mannered way for polystyrene in the span of 3 decades [9], although, a complex behaviour was observed in the visco-elastic properties of polystyrene from the temperature range of 40ºC-80ºC, which is in accordance with T Alfrey [10].